# Geometry-driven jets underlie dispersal of plants and fungi by raindrops

**Authors:** Ana-Maria Bratu[1]†, Valentin Laplaud[1]†, Antoine Garcia[1], Christophe Josserand[1], Stéphanie Drevensek[1], Camille Duprat[1], Arezki Boudaoud[1]*

[1]LadHyX, CNRS, Ecole polytechnique, Institut Polytechnique de Paris, 91128 Palaiseau Cedex, France

*Corresponding author. Email: arezki.boudaoud@polytechnique.edu

†First coauthors

**Abstract**
The impact of droplets on concave surfaces is poorly understood, although it is relevant to a mode of dispersal that has evolved independently in several species of plants and fungi. This mode relies on splash-cups, specialized organs that use raindrops to disperse reproductive units away from the parent organism. We investigated the impact of droplets on conical cavities that mimic splash-cups and we found that such impact may lead to the formation of two types of jets, which appear essential for dispersal in nature. We built a minimal kinematic model that explains jet formation, involving the motion of fluid particles along geodesics (shortest paths) on the cone surface and we predicted cone angles that optimize jet formation, consistent with the geometries of natural splash-cups.

**Main Text:**

Plants and fungi use a variety of strategies to spread their reproductive units, notably using flow of air [1], [2], [3], [4] or of water[5], [6], [7], [8]. Seeds can be carried away by flowing streams, helping them reach new environments [8]. Fungal spores can be spread from leaf to leaf by the splashing of raindrops, especially for pathogenic species, leading to plant infections and disease outbreaks [6], [9], [10]. Strikingly, several species of plants and fungi have independently evolved cup-shaped organs known as splash-cups, that use rain for dispersal [11]. The reproductive units contained in these splash-cups are carried away from the parent organism by raindrops, enabling spreading around the parental organisms and increasing the probability of effective establishment and maintenance in an already colonized habitat. Dispersal by splash-cups was first documented in natural settings [12] and later explored with a fluid mechanics perspective [13], [14], [15], [16]. Nevertheless, droplet impact on such complex geometries remains understudied. Extensive research has been devoted to droplet impact on flat surfaces, including textured surfaces, reviewed for instance in [17], [18]. Fewer studies have examined impacts on inclined [19], [20], [21], [22], [23] or non-flat surfaces[24], [25], [26], [27], [28], [29], [30]. Accordingly, the physical mechanisms underlying dispersal by splash-cups are still to be unravelled. Here, we combined experiments and models to investigate the impact of droplets on conical cavities and to make predictions about the efficiency of dispersal in the field.

To illustrate this mode of dispersal, we considered the common liverwort (*Marchantia polymorpha*, Fig. 1a), a model plant that produces vegetative clones inside splash-cups (Fig. 1b) and uses rain to disperse them [31]. The geometry of the top of splash-cups can be approximated as a conical cavity (Fig. 1c). We observed water droplets impacting Marchantia splash-cups using high-speed cameras (Fig. 1d, e). In a few milliseconds, the droplet forms an asymmetric corolla, a jet extends laterally and breaks up in smaller droplets that carry reproductive units. These steps differ significantly from the axisymmetric spreading and retraction of droplets impacting on flat surfaces, suggesting that splash-cup geometry is important for dispersal. To test this, we investigated the impact of droplets on biomimetic conical cavities obtained using 3D printing (Fig. 2e). We varied the aperture angle of the cone and the diameter of the cavity, as well as the drop diameter and the off-centering distance, defined as the lateral offset between the centers of the drop and of the cup. The impact velocity was in the range 2.9 to 3.6 m/s, with the same magnitude as the terminal velocity of droplets of the same diameter falling in air, as would be expected for raindrops.

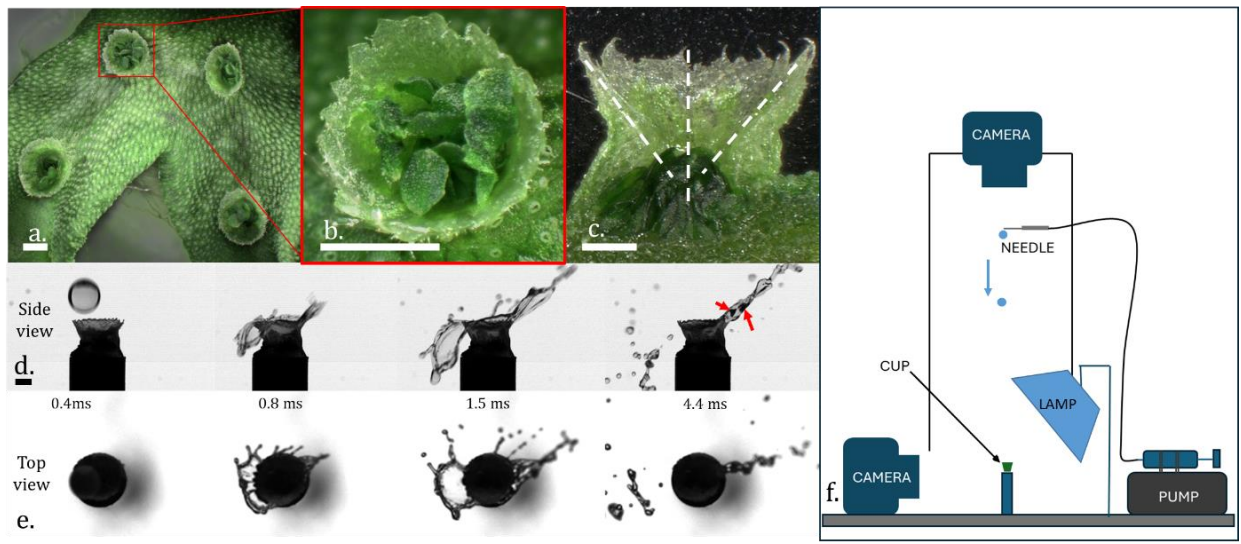

**Figure 1. Biological inspiration and experimental setup.** All scale bars represent 1 mm. **a.** Image of a *Marchantia polymorpha* thallus. **b.** Close-up of a splash-cup; the green lentil-shaped corpuscles are the reproductive units. **c.** Cross-section of a splash-cup, revealing its conical geometry. **d,e.** Visualization of a droplet (diameter: 2.00±0.01 mm) impacting the splash-cup (impact velocity: 3.07±0.01 m/s), captured from the side (d) and from the top (e), see Movie S1. The red arrows point to the reproductive units being ejected from the cup. **f.** Schematic of the experimental setup, including two high-speed cameras, a syringe pump for droplet generation via a needle, and a blue-light (470 nm) lamp to illuminate fluorescein dye added to the water.

With biomimetic cups, we observed 4 hydrodynamic behaviors shown in Figure 2 (a-d). In this figure, the drop size and impact velocity are kept constant across experiments and the control parameters are the off-centering and the cone angle. In the case of centered drop impact (Fig. 2a), a circular corolla forms, expanding to a maximal diameter before retracting forming ligaments that eventually fragment into secondary droplets very similarly to a droplet impacting a flat-ended pillar [32]. Off-centered impact leads to the formation of a jet that can be surrounded by a fluid side sheet (Fig. 1b) or can develop alone (Fig. 1c) depending on the off-centering and the amount of water that enters the cup. In steep cups, the jet is directed upward by detaching from the inner surface of the cup before reaching its edge (Fig. 1d). To capture the overall dynamics, we computed the summed intensity projections shown in Fig. 2f and 2g by integrating all video frames over time, corresponding to the events in Fig. 2c and 2d. These projections reveal the temporal evolution of the ejection angle. For blunt cups (here 45°), the jet fluctuates around the tangent to the cone and exits the cone opposite to the impact point (with respect to the cone tip). In contrast, steep cups (here 20°) can produce nearly vertical jets, which initially exit the cone on the side of the impact point.

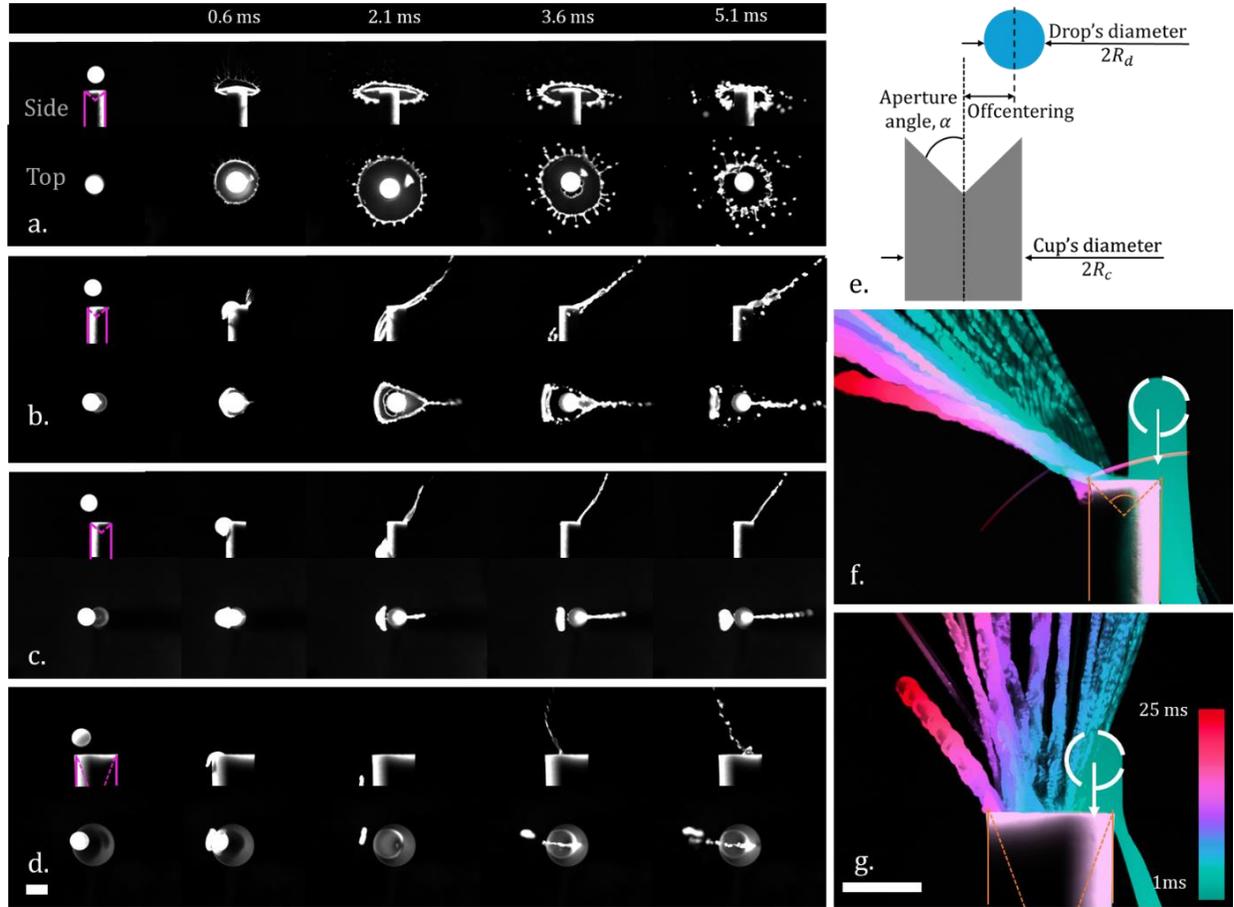

**Figure 2. Behaviors of droplets impacting concave surfaces.** In all experiments, drop diameter is 3.3 ±0.1 mm and impact velocity is 3.2±0.2 m/s; cone diameter is 3.5 mm (a,b,c,f) and 7 mm (d,g). **a.** Centered impact on a biomimetic cup with a 45° cone angle, resulting in the formation of a circular corolla that expands and then retracts, producing finger-like structures that break into secondary droplets, see Movie S2. **b.** Slightly off-centered impact on a 45° cone angle cup, leading to the formation of a lateral jet and an outward-spreading fluid sheet, see Movie S3. **c.** Off-centered impact on a 45° cone angle cup, producing a simple lateral jet, see Movie S5. **d.** Off-centered impact on a 20° cone angle cup, resulting in the formation of an upward jet, see Movie S4. **e.** Schematic of the 3D-printed biomimetic cups showing the control parameters used in the experiment. **f.** Superimposed videos showing a side view of lateral jet formation. **g.** Superimposed video frames showing a side view of upward jet formation.

For nearly centered impacts, we found that droplets produced by sheet fragmentation do not reach farther than about 5 cm from the cup. In contrast, we observed that droplets produced by jet fragmentation may reach as far as about 1 m. Therefore, jets appear as the most efficient way to disperse reproductive units. In order to quantify jet formation, we introduced a shape index that measures the shape of the sheet and jet at the time at which the sheet starts retracting (Fig. 3a.i-iii),

$$S = \frac{(A+B)/2}{C}$$

Here, A and B represent the maximum lateral extensions of sheet, while C denotes the length measured in the maximal advancing direction (i.e. the jet length). The shape index is 0 for a sim-ple jet (Fig. 3a.iii), where there is no sheet, and tends to 1 for a circular corolla (Fig. 3a.i). The index falls between 0 and 1 for intermediate cases (Fig. 3a.ii). Figure 3a presents a phase diagram of the shape index for a cup with a 45° cone angle, plotted in the parameter space defined by the normalized drop radius and normalized off-centering (both relative to the cup radius). Jet formation is more likely when the drop is smaller than the cup and its center falls outside the cup perimeter. As drop size increases, a larger off-centering is required to produce a simple lateral jet. Figure 3b shows the phase diagram for a 20° cone angle. Upward jets (highlighted in blue) occur exclusively for drops smaller than the cup and impacting inside the cup. The jetting regime delimitation shifts with cup angle, as shown in Fig. 3c (see Fig. S1 for the cases of 30° and 70°). Overall, the likelihood of simple jet formation increases when cups become steeper. This trend is further illustrated in Fig. 3d, which plots the shape index as a function of normalized off-centering for a fixed drop-to-cup radius ratio and different cone angles (see Fig. S1 for another drop-to-cup radius ratio). The impact produces a circular corolla when the drop is centered. As off-centering increases, the jet becomes increasingly pronounced, more so for a cone of steeper angle. Finally, we considered the jet velocity to assess the efficiency of the cup. In Figs. 3e and 3f, we reported vertical and horizontal components of the jet tip velocity, for the first millisecond after the impact (see 3D velocity in Fig. S1). Velocities appear to be maximal for intermediate values of off-centering, though experimental errors make it difficult to conclude. As could be expected if jets are tangential to the cone, the highest vertical velocities occur for steep cups (20°), whereas the highest horizontal velocities are reached for blunt cups (70°).

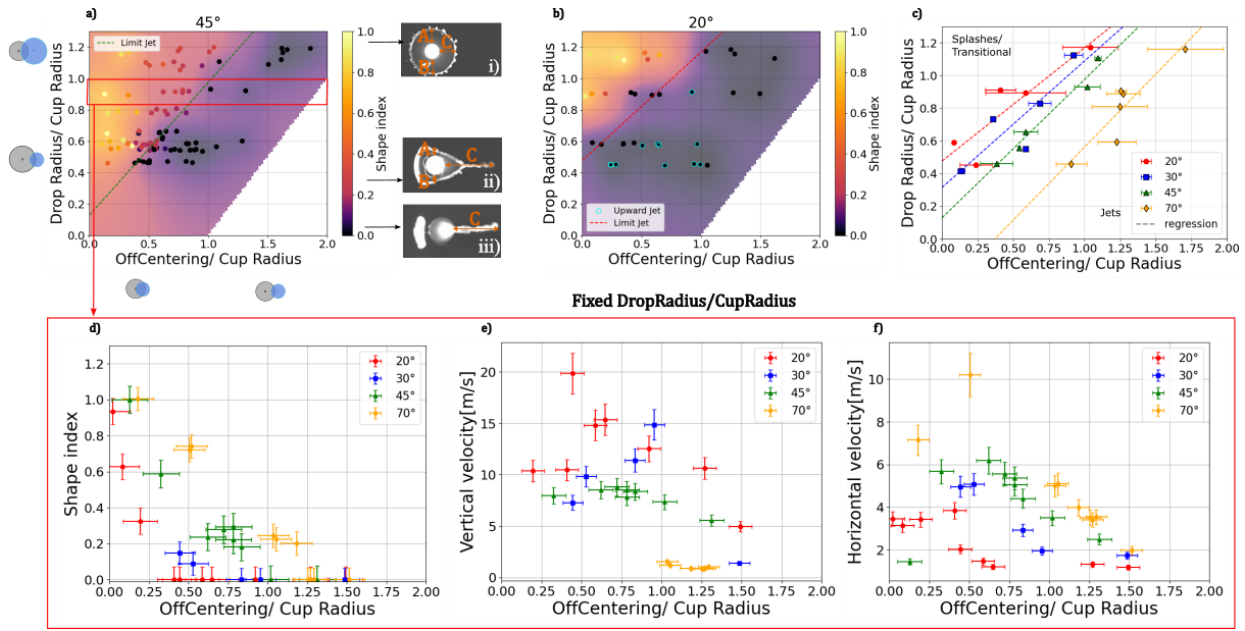

**Figure 3. Quantitative study of droplet impact on concave surfaces. a-b.** Phase diagram showing the shape index for impacts on cups with cone angles of 45° (a) and 20° (b). Experimental data are shown as dots. The background is included as a visual aid for the reader; it represents an interpolation of shape index values in the phase space, generated through inverse distance weighting (exponent = 2). The green and red dashed lines indicate the boundary between jetting and splashing/transitional behaviors; they were obtained by linear regression through the leftmost points (for a fixed drop-to-cup radius ratio) with a shape index equal to zero. Experiments where an upward jet was observed are circled in blue. **c.** Jetting regime boundaries for four different cone angles. **d–f.** Shape index (d), vertical (e), and horizontal (f) jet tip velocities vs. normalized off-centering for all four cone angles at fixed ratio of drop radius to cup radius.

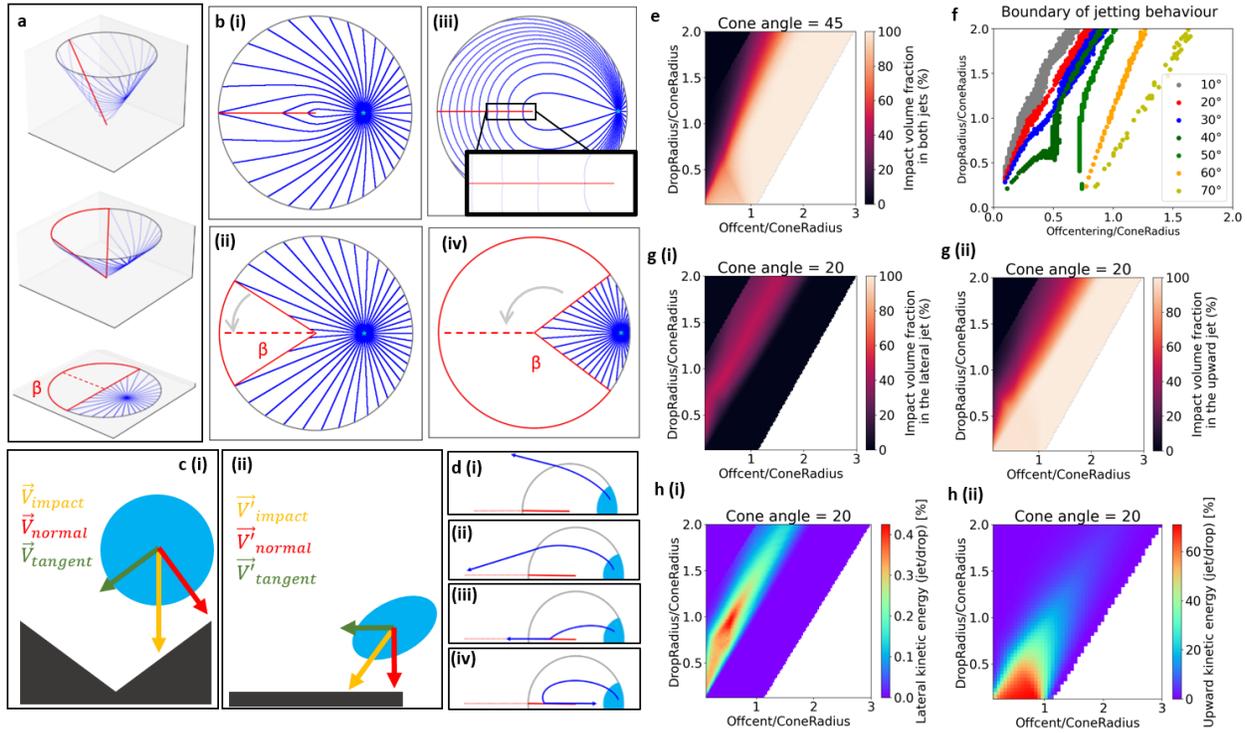

**Figure 4. Kinematic model for drop impact. a.** Unfolding of a cone into a disk sector (top to bottom). The blue lines are geodesics (shorter paths on the surface of the cone) originating from a single point. The removed sector (red) angle β is directly related to the cone internal angle. Here, the cone angle has the value 30° (corresponding to β=180°), which is the threshold that separates the two cases of panel (b). **b.** Top view of the cone **(i, iii)** and unfolded disk sector **(ii, iv)** configurations for cones of angles α = 55° and 15° respectively. The *cone configuration* shows symmetric pairs of geodesics meeting at the removed sector (red). In the *disk configuration*, folding is depicted by a gray arrow. **c.** Side views of the decomposition of the impact velocity into normal and tangent components, in *cone configuration* **(i)** and *disk configuration* **(ii)**. The ellipsoid shape of the drop is not accurate and represents the deformed 3D shape of the drop in the *disk configuration*. **d.** Schematics of the different trajectories for a fluid particle. **(i)** The fluid particle reaches the edge of the cone, contributing to the sheet. **(ii)** The fluid particle reaches the edge of the cone and joins the lateral jet outside the cone. **(iii)** The fluid particle collides with its symmetric counterpart and forms the lateral jet. **(iv)** For cone angles < 30° the fluid particle trajectory can be bent enough so that it collides with its symmetric counterpart with a reentrant angle, contributing to an upward jet. **e.** Volume fraction in the jet (proportion of the volume of the incoming drop that goes into the jet) for a cone of 45° angle **f.** Evolution with the cone angle of the position of the boundary of the jetting regime (defined as volume fraction in the range 94-96 %). **g.** Volume fractions in the lateral jet **(i)** and in the upward jet **(ii)** for a cone of angle 20°. **h.** Kinetic energy available for dispersal in the lateral **(i)** and upward **(ii)** jets, for a cone of angle 20°.

In order to get a better understanding of jet formation, we built a minimal model of impact, aiming at a semi-quantitative agreement with experimental results. We first assessed the relative weight of forces at play by considering relevant dimensionless numbers. We considered typical experimental values: a drop made of water (density $\rho = 1000$ kg.m$^{-3}$, kinematic viscosity $v = 1.0 \times 10^{-6}$ m$^2$/s, surface tension $\gamma = 0.07$ N/m) falling at velocity $V = 3.5$ m/s on a cone of diameter $2R_c = 3$ mm, due to gravity ($g = 9.8$ m.s$^{-2}$). The Reynolds number, which compares inertia to viscous forces, takes the value $V\,2R_c / v \sim 10000$; the Weber number, which compares inertia to capillary forces, is $\rho\, V^2\, 2R_c\, /\, \gamma \sim 500$; the Froude number, which compares inertia to gravity, is about $V\, /\, (g\, 2R_c)^{1/2} \sim 20$. All dimensionless numbers are large, so that the dynamics is dominated by inertia. In impact on flat surfaces, inertia leads to fluid particles radiating from the center of impact along lines, i.e. along shortest paths on a plane. The core of our model is to assume that fluid particles follow geodesics (shortest paths) on the cone. Geodesics on a cone can be readily constructed by 'unfolding' the surface of the 3D cone into a 2D disk sector (Fig. 4a); the angle of the sector ($\beta$) depends on the cone angle ($\alpha$) via $\beta = 2\pi\, (1 - \sin \alpha)$. We computed analytically the forward and reverse transformations between the 2D projections of the folded (*cone configuration*) and unfolded (*disk configuration*), see Methods and Fig. 4b (more examples in Fig. S2), yielding the geodesics shown in Fig. 4a,b. A striking feature is that pairs of geodesics emanating from the impact point meet on a disk radius opposite this point; fluid particles meeting there would be deflected, merge and start moving along the radius, explaining jet formation at that radius. We proceeded similarly to drop impact on an inclined surface [33] (Fig. 4c) that has a size comparable to the drop [32] and we defined fluid particles trajectories and velocities on the disk surface (see Methods). Fluid trajectories may exit the cone without meeting, contributing to the sheet (Fig. 4d-i), they may meet with acute angles outside (Fig. 4d-iii) or inside (Fig. 4d-ii) the cone, contributing to the lateral jet, or they may meet with reentrant angles (Fig. 4d-iv), contributing to the upward jet.

Based on the model, we computed the proportion of the drop that feeds each type of jet, which we call the jet volume fraction. Upward jets only occur for cone angles smaller than 30°. Fig. 4e shows the phase diagram of the lateral jet for a cone angle of 45°, to be compared to the shape index shown in Fig. 3a. A simple jet, in which all the water that falls into the cup is directed into a jet, corresponds to a fraction of 100% and a shape index of 0, whereas a corolla means a fraction of 0% and a shape index of 1. Fig. 4f shows the prediction of the boundary between jetting and splashing/transitional behaviors, which semi-quantitatively recapitulates the observations of Fig. 3c. We note that the boundary is curved, suggesting that the fit to a line in Fig. 3c is an oversimplification. Figs. 4g,h show volume fraction and available related kinetic energy (see Methods) for the lateral and upward jets for a cone angle of 20°. Fig. 4g can also be semi-quantitatively compared to Fig. 3b. The available kinetic energy is higher in regions of higher jet volume fraction but does not completely correlate with it, due to variations in jet velocity with cone angle and off-centering. Actually, fluid particles meeting at small angles are less slowed down than when meeting upfront, which might explain the variations of jet velocity observed in experiments (Fig. S1d).

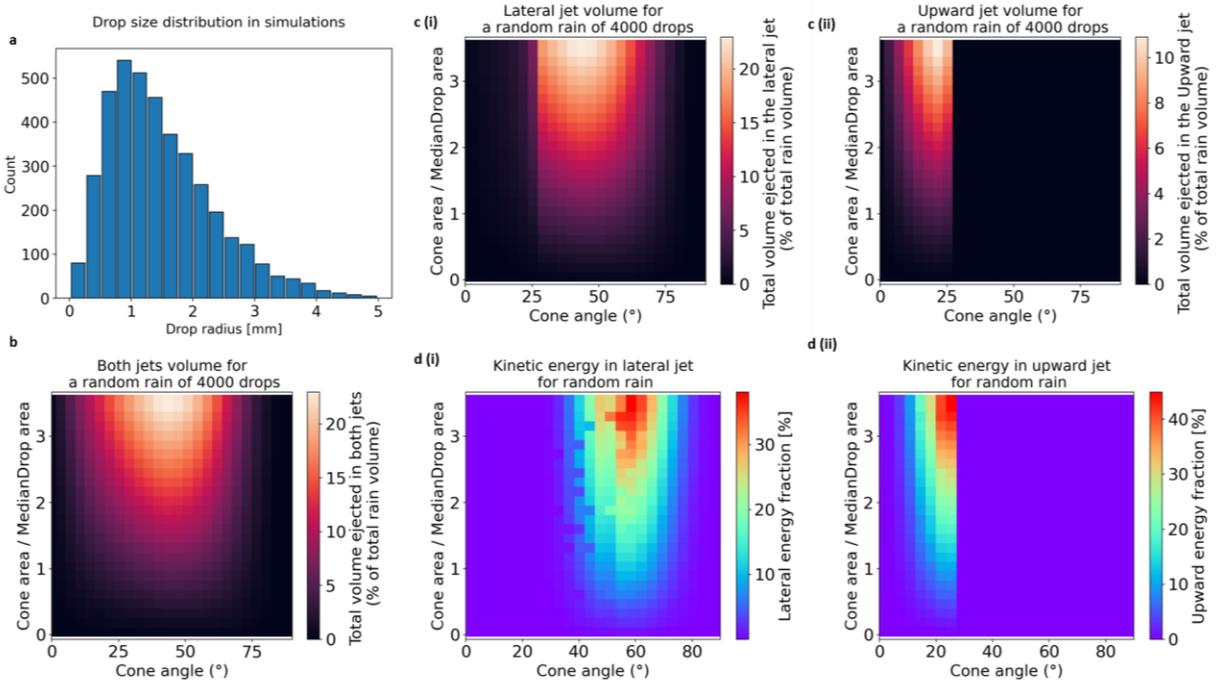

**Figure 5. Cup shape optimization for random rain**. **a.** Example of drop radius distribution in a simulated random rain. **b.** The relative volume ejected by a cone in either type of jets. **c.** Volume ejected in the lateral (i) and upward (ii) jets. **d.** Proportion of the rain kinetic energy that is available for the lateral (i) and upward (ii) jets.

Having established the model, we sought whether natural cup shapes are optimal with respect to dispersal. We modelled a random rain and quantified water ejected in jets, because only jets are efficient for dispersal. The 'random rain' is generated as a collection of drops with different sizes (using a truncated gamma distribution, Fig. 5a) and their associated terminal velocities in air. To obtain a homogeneous 2-dimensional distribution of drop impact points, we considered an axisymmetric geometry and the appropriate distribution for the distance of impact points to cup center, which vanishes at the center and is maximal at the periphery (see Methods). We computed for each cone the total rain volume fraction that is converted in jets as well as the kinetic energy available for dispersal relative to the total amount of kinetic energy of the rain. Because we consider optimization in a given environment where rain properties, and so median drop size, are given, we plotted our results as function of cone area, which is a proxy for the quantity of biological matter used to build the cup, normalized by median drop area. Fig. 5b shows the total volume fraction going in jets. A cone with larger area always allows more rain volume to be in the jets, as could be expected with a larger area to intercept the rain. For the lateral jet (Fig. 5c-i), the cones ejecting more volume in jets are those with an angle around 45°. For the upward jet (forming only for cone angles below 30°, Fig. 5c-i), the maximum of volume in jets is achieved for cone angles close to 25°. In both cases, maxima exist due to the balance between two effects: with a larger angle the projected area is larger (for a fixed cone area) while with a smaller angle more trajectories are focused and form jets. Fig. 5d shows the kinetic energy associated to jets. For both types of jets, the cone angle that maximizes kinetic energy is larger than the angle for maximizing the volume. The upward jets carry more kinetic energy at the critical cone angle of their formation, 30°. For the lateral jet the energy maximum is reached for a cone angle of about 60°. We discuss hereafter the relevance of these optimal configurations to natural settings.

Our results reveal two distinct dynamical regimes when a drop impacts a conical cavity, and delineate their boundaries in the morphological parameter space. In the centered-impact case on conical cups, a circular corolla forms, whereas off-center impacts give rise to jet formation, potentially combined with fluid sheets at intermediate off-centering. We found that only jets are efficient to carry reproductive units away from the parent organism. We identified two qualitatively different jetting behaviors. Whatever the cup angle, lateral jets are ejected around the tangent to the cup surface. For steep cups (cone angle ≤30°), upward jets detach from the cup inner surface and leave the cup close to the vertical. Based on a kinematic model of jet formation, we predicted cone angles that could be optimal for dispersal, given a distribution of raindrop sizes. We used two criteria for optimality: (i) maximizing the ejected volume of water means forming as much jet as possible, potentially maximizing the number of reproductive units carried away by jets; (ii) maximizing available kinetic energy means increasing jet velocity, potentially increasing the reach of reproductive units carried away. We found two optima. The optimum for lateral jets is achieved for cone angles of ~45° and ~60° according to criterion (i) and (ii), respectively. These optima can be compared to the angle of splash-cups in flowering plants, which ranges from 37 to 63° [14], or in *Marchantia polymorpha*, which is around 35° (Fig. 1c). The range of natural angles is rather broad. Nevertheless, lateral jets seem suited to plants because they favor horizontal spreading, consistent with observed tangential jets in *M. polymorpha* (Fig. 1d). The optimum for upward jets is achieved for cone angles of ~25° and ~30° according to criterion (i) and (ii), respectively. These optima can be compared to the steep angles of splash-cups in fungi, which are around 20° for bird nest fungus, for instance [13]. Upward jets seem suited to these species in which, after dispersal, reproductive bodies typically attach to plants above the fungal splash-cup. Indeed, experiments on bird nest fungus show nearly vertical jets that do not touch the walls of the splash-cup [13]. Altogether, we have described a geometry-based mechanism for jet formation, which is relevant to dispersal of plants and fungi by rain and partially explains the geometry of their splash-cups.

**Acknowledgments:** We thank Thibault Perrin and Cyril Hasson for their help with 3D printing and Catherine Reeb for discussions on modes of dispersal.

**Funding:**

Agence Nationale de la Recherche grant ANR-23-CE13-0035-01

This project has received financial support from the CNRS through the MITI interdisciplinary programs.

**Author contributions:**

Conceptualization: CJ, SD, CD, AB

Methodology: AMB, VL, AG, CD, AB

Investigation: AMB, VL

Visualization: AMB, VL

Funding acquisition: CJ, SD, CD, AB

Project administration: SD, AB

Supervision: SD, CD, AB

Writing – original draft: AMB, VL, AB

Writing – review & editing: AMB, VL, SD, CD, AB

**Competing interests:** Authors declare that they have no competing interests.

**Data and materials availability:** All data and code are available upon request.

# Supplementary Materials for

# Geometry-driven jets underlie dispersal of plants and fungi by raindrops

Ana-Maria Bratu[1]†, Valentin Laplaud[1]†, Antoine Garcia[1], Christophe Josserand[1], Stéphanie Drevensek[1], Camille Duprat[1], Arezki Boudaoud[1]*

*Corresponding author. Email: arezki.boudaoud@polytechnique.edu
†First coauthors

**The PDF file includes:**

    Figs. S1 to S2
    Materials and Methods
    References

**Other Supplementary Materials for this manuscript include the following:**

    Movies S1 to S10

**Fig. S1.**

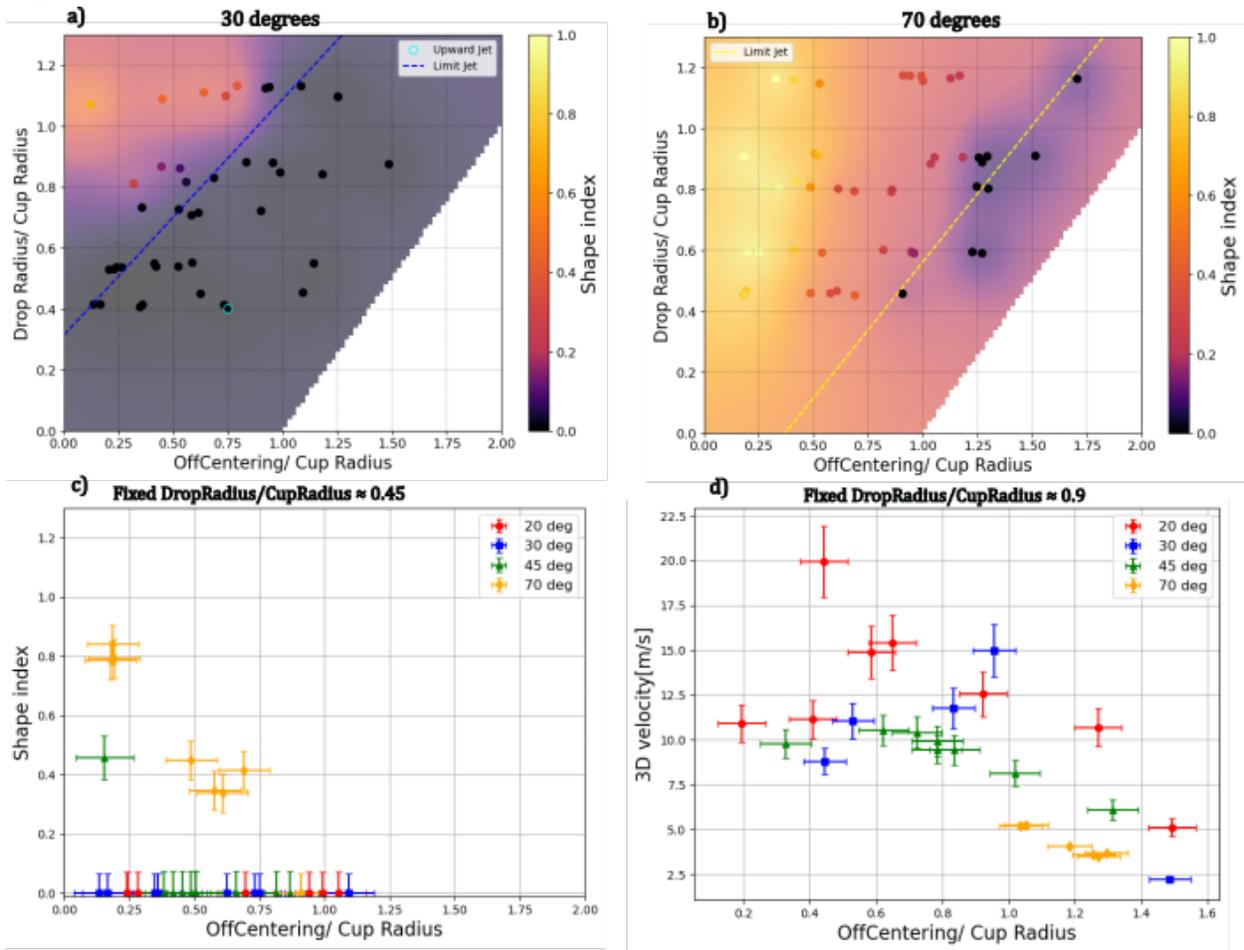

**Impact of droplets on conical cavities. a.** Phase diagram showing the shape index for impacts on a cup with a 30° cone angle, the experiment where an upward jet was observed is circled in blue. **b.** Phase diagram of the top view shape index for a 70° cone angle cup. **c,d.** Shape index (c) and norm of the velocity of the jet tip (d) vs. normalized off-centering for all four cone angles at fixed DropRadius to CupRadius ratio.

**Fig. S2.**

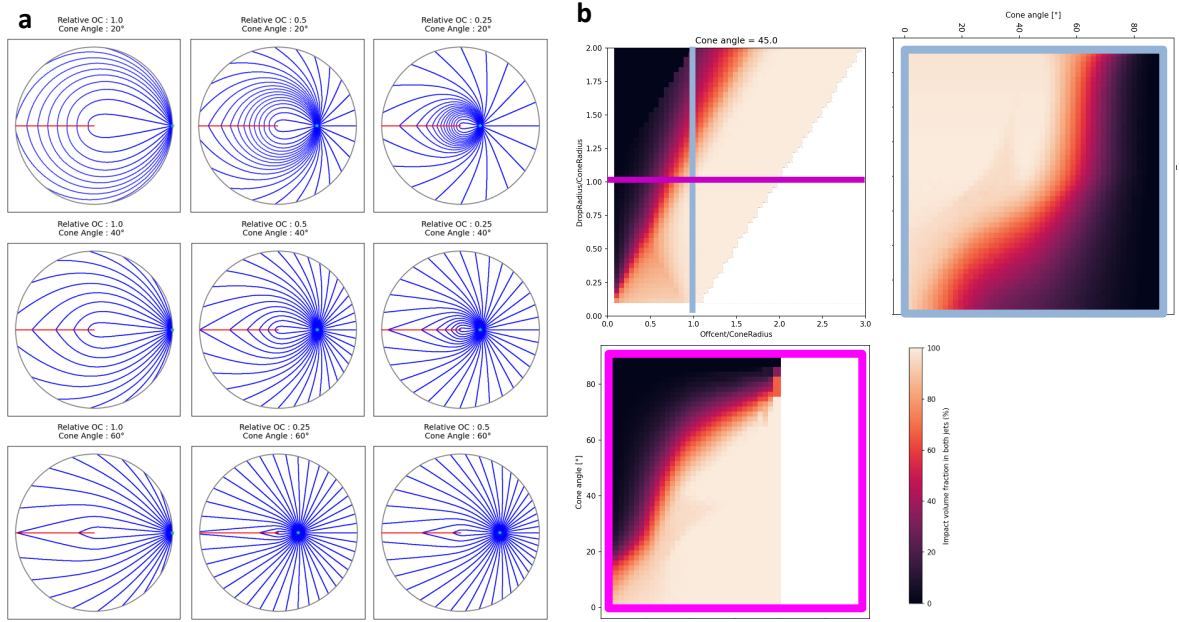

**Kinematic model of droplet impact. a.** Illustration of geodesics on the cone's surface for different impact positions and cone angles. **b.** Projected views of the 3D parameter space for the total volume fraction of water going into jets.

**Movie S1.**

Side and top view of a water droplet falling on a Marchantia splash-cup, corresponding to Fig. 1d,e.

**Movie S2.**

Side and top view of a droplet falling on a conical cavity of angle 45°, corresponding to Fig. 1a.

**Movie S3.**

Side and top view of a droplet falling on a conical cavity of angle 45°, corresponding to Fig. 1b.

**Movie S4.**

Side and top view of a droplet falling on a conical cavity of angle 45°, corresponding to Fig. 1c.

**Movie S5.**

Side and top view of a droplet falling on a conical cavity of angle 20°, corresponding to Fig. 1d.

# Materials and methods

## 1 Experimental details

### 1.1 Experiments

Biomimetic cups were 3D-printed from PlasCLEAR V2 resin using an Asiga Max UV385 printer. The measured static contact angle of water on the resin surface was 80°. All experiments were performed on dry target surfaces ; each cup was dried prior to droplet impact using compressed air. Droplets were generated using custom-made glass needles of inner diameter in the range 100-200 $\mu$m, to enable the production of droplets with diameters below the capillary length. Some of the needles were coated with a superhydrophobic 3M™ Novec™ 1720 Electronic Grade Coating layer to produce smaller droplet. To enhance optical contrast, fluorescein (46955-1G-F Sigma Aldrich) was added to the water at a concentration of 0.2 mM that did not significantly affect the properties of water. Illumination was provided by a blue light source (THORLABS SOLIS-470C). The dynamics was monitored at 10000 frames per second using two high-speed cameras : a Photron Mini AX200 (1024x672) positioned for top view imaging and a Photron Mini UX100 (1280x480) for side view acquisition.

### 1.2 Quantifications

Image processing methods were internally developed in Python, using custom scripts built using SciPy and scikit-image libraries. The drop and cup radii were obtained from top-view images by fitting circles to their outlines. The off-centering was then calculated as the distance between the centers of these two circles.The shape factor was obtained from segmented top images by measuring the extent of the sheet/jet along the maximal advancing direction at the moment of sheet retraction. The side sheet extension was measured along the direction perpendicular to this maximal advancing direction at the same moment. Velocities were determined from image sequences. The vertical velocity was obtained from side-view images by segmenting the jet and tracking its tip in the vertical direction. The time evolution of this position was fitted with a straight line, and the slope provided the velocity (assuming that, at early times, the velocity is approximately constant). The horizontal velocity was deduced in the same way from top-view images by tracking the jet tip along the maximal advancing direction.

## 2 Model details

Our objective is to propose a minimal model of jet formation. We are interested in quantifying the efficiency of dispersal by computing the amount of water projected in the jet, and by estimating the available kinetic energy and the range of dispersal.

## 2.1 Problem description

### 2.1.1 Geometry

We consider a drop of radius $R_d$ falling on a cone of radius $R_c$ and opening angle $\alpha$ (Fig. 1a). Their centers are separated by the offcentering $r$ such that: $0 < r < R_c + R_d$ (Fig. 1b).

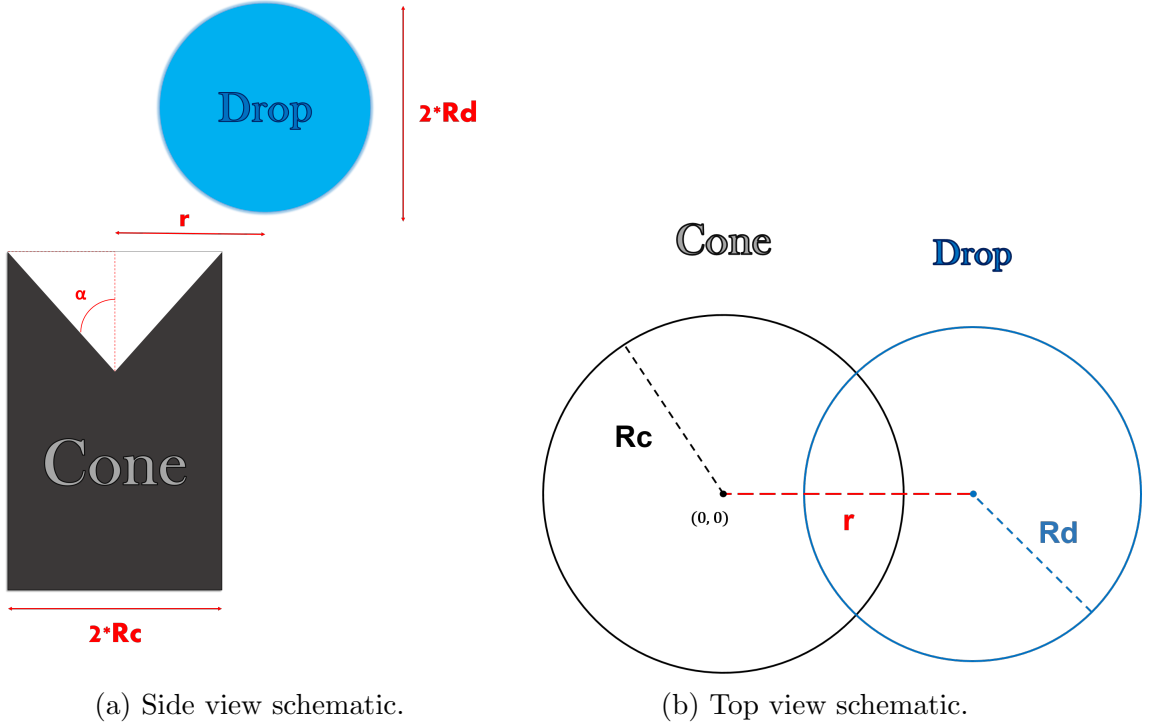

(a) Side view schematic.  (b) Top view schematic.

Figure 1: Geometry of the problem

### 2.1.2 Dimensionless numbers

We consider raindrops falling onto mm-sized cones. We used the following parameters: Impact speed $V = 5\,m.s^{-1}$, Cone diameter $2R_c = 3\,mm = 0.003\,m$, Water density $\rho = 1000\,kg.m^{-3}$, Water viscosity $\mu = 0.001\,Pa.s^{-1}$, Water surface tension $\gamma = 0.07\,N.m^{-1}$, and the gravitational constant $g = 9.8\,m.s^{-2}$. We can thus compute a few dimensionless numbers of fluid mechanics relevant for our problem: Reynolds (**Re**, viscosity vs. inertia), Weber (**We**, capilarity vs. inertia), and Froude (**Fr**, gravity vs. inertia)

$$\mathbf{Re} = \rho 2 R_c V / \mu \approx 10000.$$
$$\mathbf{We} = \rho V^2 2 R_c / \gamma \approx 500,$$
$$\mathbf{Fr} = V / \sqrt{g 2 R_c} \approx 20.$$

From these we can see that the impact process is dominated by inertia. We will thus start by devlopping an inertial, kinematic model to explain the formation of the jet. We will investigate fluid particles trajectories at the surface of the cone, based on geodesics of the cone.

### 2.1.3 Trajectories at the surface of the cone

In this model we consider the curved surface of the cone as resulting from the folding of a flat surface. Indeed a cone of angle $\alpha$ can be folded from a disk with a removed sector of angle $\beta$ (Fig. 2a):

$$\beta = 2\pi(1 - \sin(\alpha))$$

We will then resolve the impact and figure out the fluid particles trajectories on the flat partial disk and transform them back to the cone surface. We will also consider the impact to be quasi-instantaneous (impact time $\ll$ sheet expansion time) as was done in [Villermaux and Bossa, 2011], and the flow to be 2D at the surface of the cone. We always orient the $(\vec{x}, \vec{y})$ axes such that the drop center is on the $y$ axis to get $x_c = r$ and $y_c = 0$. Finally since the cone shape is fixed for a specific $\beta$ angle the $z$ coordinate of a point $M(x, y, z)$ at the surface of it is constrained by its $(x, y)$ coordinates and we can thus work with the 2D projection of the cone (Fig. 2b).

The direct transformation of the plane to go from partial disk to projected cone is simple to express in polar coordinates centred on the cone (or disk) center. With $(r, 0)$ the coordinates of the drop impact point and $(R'_M, \theta'_M)$ the coordinates of a point $(M')$ in the *disk configuration* we have the point $(M)$ in the *cone configuration* described by

$$R_M = R'_M \sin(\alpha), \quad \theta_M = \frac{\theta'_M}{\sin(\alpha)} \tag{1}$$

This transformation equation is illustrated in Fig.2b with dotted geodesics drawn in the *disk configuration* (left) and each dot displaced in the *cone configuration* (right) using (1).

Additionally to transforming point positions, we also need to know how to transport the velocities between configurations. We use the velocity decompositions in the Cartesian plane $(\vec{x}, \vec{y})$: $\vec{V_M} = V_x \vec{x} + V_y \vec{y}$ and $\vec{V_{M'}} = V'_x \vec{x} + V'_y \vec{y}$. In the *cone configuration* we have:

$$\begin{aligned} V_x &= \dot{x} & V_y &= \dot{y} \\ x &= R\cos\theta & y &= R\sin\theta \\ V_x &= \dot{R}\cos\theta - \dot{\theta}R\sin\theta & V_y &= \dot{R}\sin\theta + \dot{\theta}R\cos\theta \end{aligned}$$

Similarly for the *disk configuration* we have (using (1)):

$$V'_x = \dot{x}' = \dot{R}'\cos\theta' - \dot{\theta}'R'\sin\theta' = \frac{\dot{R}}{\sin\alpha}\cos(\theta\sin\alpha) - \dot{\theta}R\sin(\theta\sin\alpha)$$

$$V'_y = \dot{y}' = \frac{\dot{R}}{\sin\alpha}\sin(\theta\sin\alpha) + \dot{\theta}R\cos(\theta\sin\alpha)$$

Thus we can express the velocities in the *disk configuration* as:

$$V'_x = V_x \left[\frac{\cos\theta\cos(\theta\sin\alpha)}{\sin\alpha} + \sin\theta\sin(\theta\sin\alpha)\right] + V_y \left[\frac{\sin\theta\cos(\theta\sin\alpha)}{\sin\alpha} - \cos\theta\sin(\theta\sin\alpha)\right]$$

$$V'_y = V_x \left[\frac{\cos\theta\sin(\theta\sin\alpha)}{\sin\alpha} - \sin\theta\cos(\theta\sin\alpha)\right] + V_y \left[\frac{\sin\theta\sin(\theta\sin\alpha)}{\sin\alpha} + \cos\theta\cos(\theta\sin\alpha)\right]$$

$$\tag{2}$$

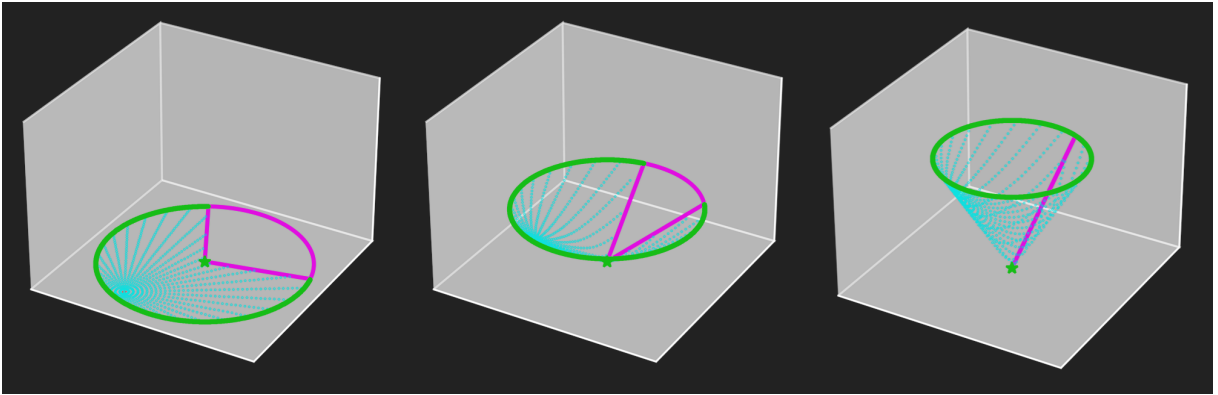

(a) Cone folding in 3D.

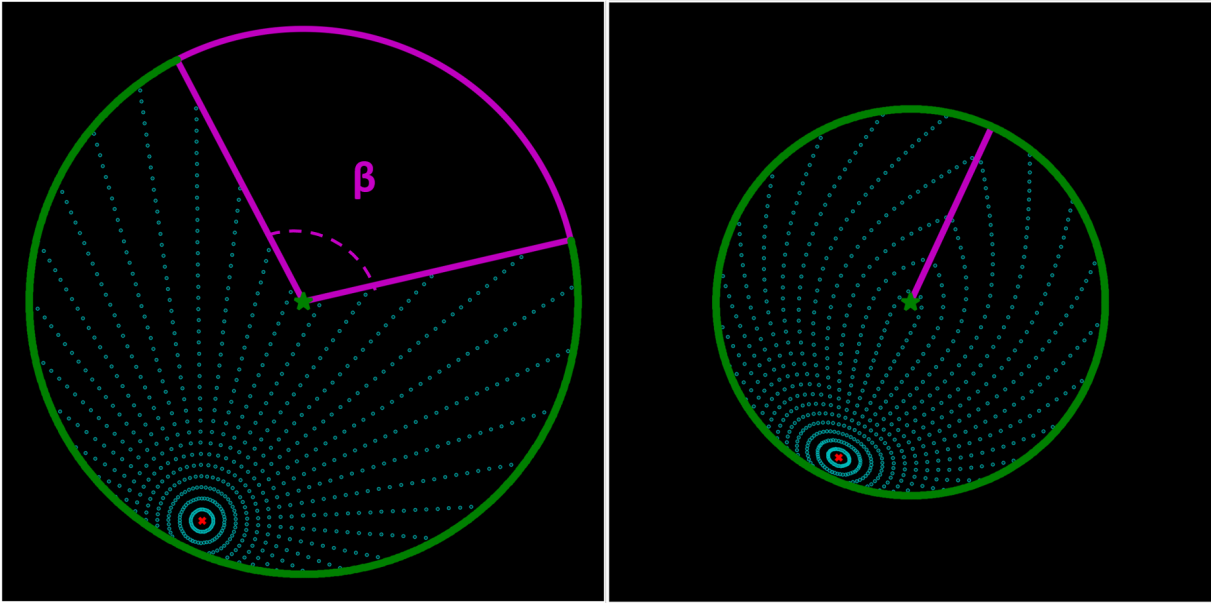

(b) 2D projection.

Figure 2: Determination of trajectories on the surface of a cone by folding a disk. In this instance radial trajectories radiating from a single point are displayed in light blue at the surface of the disk/cone.

### 2.1.4 Drop and cone of similar size

The representation used in figure 2 would correspond to a point drop impact where all the trajectories are considered radial. In the case that we want to describe the sizes of the drop and the cone are similar and there is many cases where only a fraction of the drop impacting the cone. The first step to obtain the surface trajectories is to determine the shape of the impacting area in the *disk configuration*. Then for any point in this impact area we need to assign an initial surface velocity for the fluid particle impacting that point. We then consider the initial velocity of the fluid particle to be constant as in a classical drop impact. Indeed in the case of a classical drop impact [Josserand and Thoroddsen, 2016] we know that the eulerian velocity field is:

$$u(r,t) \approx \frac{r}{t}$$

This corresponds to a constant Lagrangian velocity for each fluid particles:

$$u(t+dt) = \frac{r(t+dt)}{t+dt} = \frac{r(t)+u(t)dt}{t+dt} = \frac{r(t)+\frac{r(t)}{t}dt}{t+dt} = \frac{r(t)}{t} = u(t)$$

This mean that fluid particles will have constant velocity in the *disk configuration*.

**Drop shape in disk configuration** In the *cone configuration* the 2-D projection of the drop is an off-centred disk described by:

$$(x-x_c)^2 + (y-y_c)^2 = R_d^2$$

with $(x_c, y_c)$ the coordinates of the center. In polar coordinates centred on the cone we get for the projected drop disk equation:

$$(R\cos(\theta) - r)^2 + R^2 \sin(\theta)^2 = R_d^2 \tag{3}$$

If we use the projection transformation defined in (1) we get

$$(R'\sin(\alpha)\cos(\frac{\theta'}{\sin(\alpha)}) - r)^2 + R'^2 \sin(\alpha)^2 \sin(\frac{\theta'}{\sin(\alpha)})^2 = R_d^2$$

which we can rearrange as a $2^{nd}$ degree equation whose solutions are:

$$R' = \frac{r\cos\left(\frac{\theta'}{\sin(\alpha)}\right) \pm \sqrt{R_d^2 - r^2 \sin\left(\frac{\theta'}{\sin(\alpha)}\right)^2}}{\sin(\alpha)}$$

If $R_d > r$ the shape is described by the '+' solution and in our case, we need to limit theta to $[-\pi + \beta, \pi - \beta]$ to account for the removed sector (Fig. 3a).
If $R_d < r$ the shape is described by both solutions (Fig. 3b) with

$$|\theta'| < \sin(\alpha)\arcsin\left(\frac{R_d}{r}\right)$$

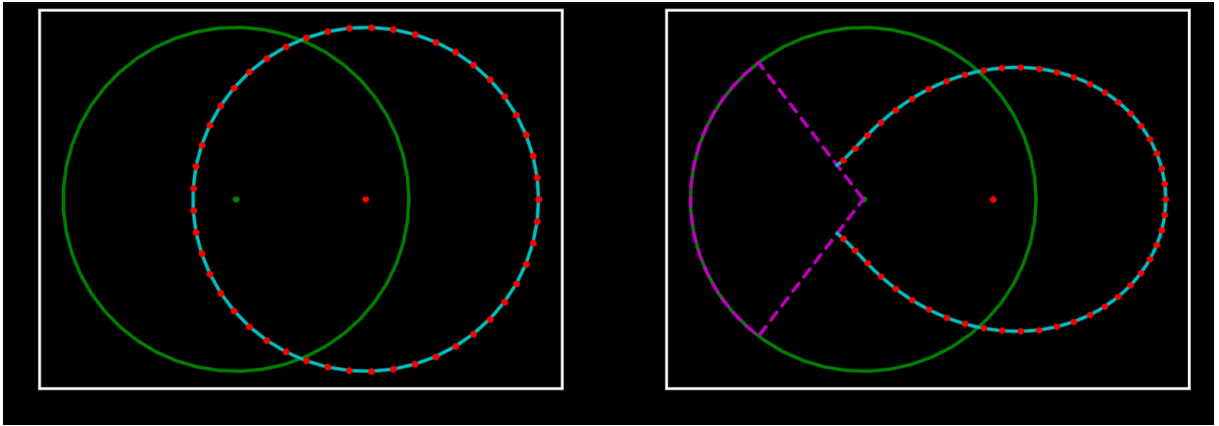

(a) Case $R_d > r$ for *cone configuration* (left) and *disk configuration* (right)

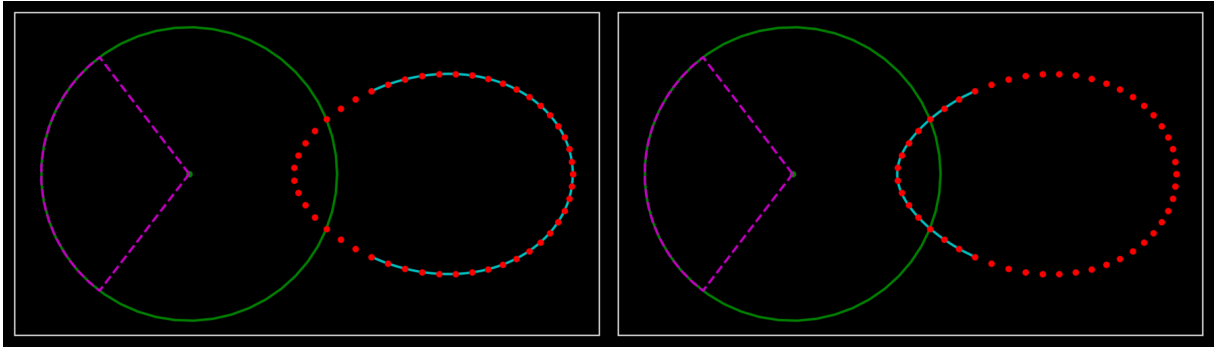

(b) Case $R_d < r$ for *disk configuration* only

Figure 3: Drop shape computation in disk configuration. On each subfigure the green disk representes the cone, the red points are placed on the drop disk in *cone configuration* and are individually transformed using eq. (1), and the cyan lines are drawn from eq. (3) and the corresponding constraint on $\theta'$.

**Trajectories in disk configuration** To properly model rectilinear trajectories in this situation we only need two pieces of information: starting points and initial velocities. Since we consider the situation to be only inertial and no interaction with the substrate, the fluid particles directions and velocities will not change in time. The starting points of trajectories are all points on the surface of the cone where water is impacting (e.g. M3 in Fig. 4). In the 2D projection *cone configuration* they are in the intersection of two disk and thus their coordinates $(x, y)$ verify

$$x^2 + y^2 \leq R_c^2 \qquad (4)$$

and

$$(x - r)^2 + y^2 \leq R_d^2 \qquad (5)$$

Regarding the initial velocities, we need both the direction and the value of the velocity, for each starting point. We want those parameters in the disk configuration, as it is the one with a flat surface on which the trajectories will be lines. First we will decompose the impacting velocity on the cone in its normal and tangential components with regard to the cone's surface. This decomposition is difficult to visualise in 3D with the projection directions depending on the angle. However with our set of hypothesis of negligible viscosity and surface tension we can consider the drop as an ensemble of fluid particles that only interact sterically with each other. In that case, to each impacting point on the cone surface there corresponds a column of fluid particle $i$ that will impact this point with a speed $\vec{V_0} = V_{norm}\vec{i1} + V_{tan}\vec{i2}$ where the decomposition base($\vec{i1}, \vec{i2}$) depends on the position (Fig. 5a). In the disk configuration, the 3D shape of the drop is not straightforward due to the 3D space deformation that is much more complex that the projected transformation explicited in eq. (1). Nevertheless we can take our previous observation in the *cone configuration* and know that every impact point has a certain amount of fluid particles reaching it with a velocity $\vec{Vl_0} = V_{norm}\vec{z} - V_{tan}\vec{r}$ in polar coordinates (Fig. 5b). At order 0 we can consider that there is an 'drop impact'-like contact with the flat surface with a velocity $V_{norm}$ resulting in a classical splash that is advected by a velocity $V_{tan}$. It is important to note that the advection velocity in the *disk configuration* is centripetal and thus concentrating the flow. We will see later how to deal with this and implement order 0 incompressibility of the fluid. To get the final velocity vector at each impacting point we will add the velocity vectors spawning from both the impact and the advection.

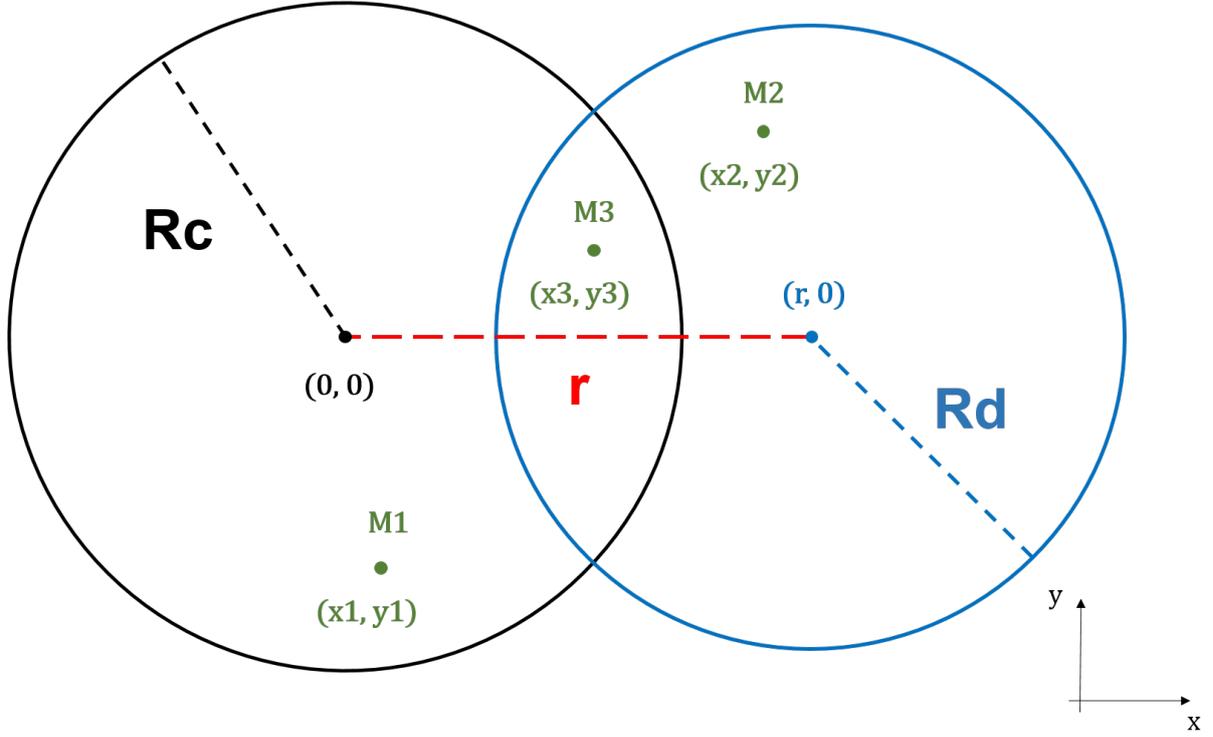

Figure 4: Illustration of the condition for impacting point. M1 and M2 are not impacting points, M3 is.

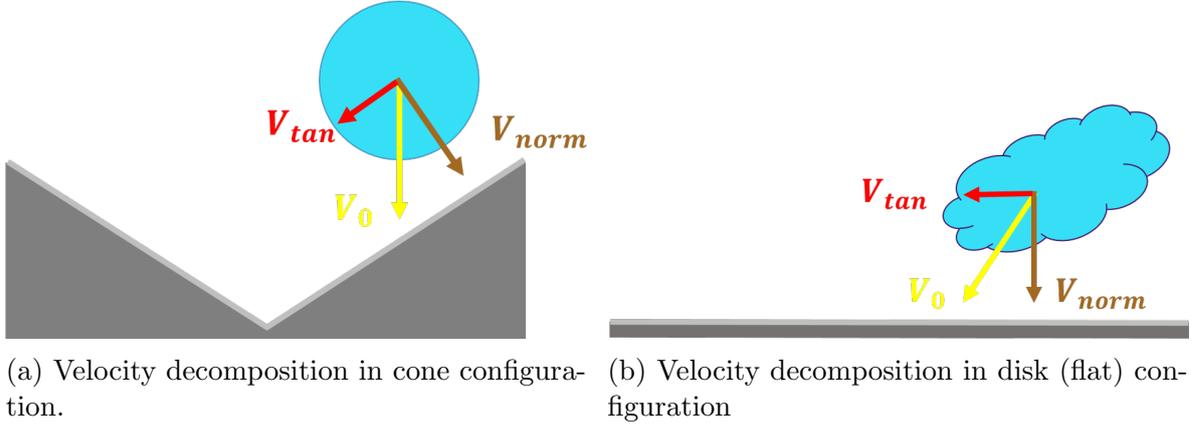

(a) Velocity decomposition in cone configuration.

(b) Velocity decomposition in disk (flat) configuration

Figure 5: Impact velocity decomposition.

**Fluid velocity associated to impact velocity** $V_{norm}$. We will first deal with the velocities orientation from the normal impact at $\vec{V}_{norm}$. In the case of a spherical drop impacting a flat surface we know that the trajectories are radial from the impact center. In addition we know that there is a stagnation point at the center of the drop and that the velocity at the front is the same as the impact velocity $V_{norm}$ [Josserand and Thoroddsen, 2016]. Within the impact disk the radial velocity norm evolves as

$$u(r_d, t = 0) = \frac{r_d}{R_d} V_{norm} \qquad (6)$$

where $R_d$ is the radius of the drop and $r_d$ is the distance to the center of the drop.

Here our drop sometimes only partially impacts, and the drop shape projected in the *disk configuration* is not a disk. To adapt the classical drop impact model to our deformed drop geometry we considered that the orientation of each fluid particles trajectories will be defined by the local thickness gradient of the drop. If transposed for a spherical drop impact this view fits with radial trajectories, and it in our model it is a weak analogy to a pressure gradient. For the norm of the velocity we again go back to the case of a full drop impacting a flat surface. Since we only have a partial impact of the drop we will use formula (6) and replace $r_d$ and $R_d$ by $r_i$ and $R_i$. $r_i$ is the distance between the impact point considered and the point of maximum drop thickness (analog to the drop center in a normal impact). For $R_i$ we take the distance between the maximum thickness (the 'center') and the furthest edge of the impacting region. The obtained velocities in two different off-centring cases are shown in figure 6.

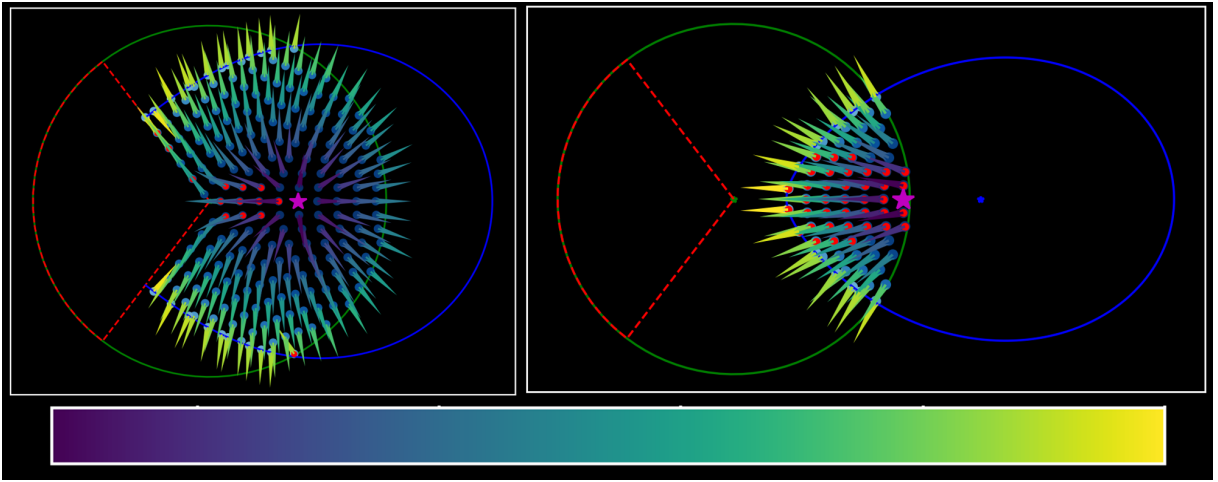

Figure 6: Velocity from normal impact for two cases. The left case is with $\frac{r}{R_d} = 0.5$ and the right case with $\frac{r}{R_d} = 1.4$.

**Fluid velocity associated to advection velocity $V_{tan}$.** We must now determine the velocity component due to the advection tangential to the surface. As stated earlier the projection $V_{tan}$ is centripetal (Fig. 7). This 2D flow is obviously not compatible with water being incompressible. One order 0 solution to get an incompressible flow is to compute the mean flow and assign its value to all fluid particles, however this cannot be done directly in the *disk configuration* as a part of space is missing (the removed sector). To overcome this difficulty with consider the tangential flow projected in the *cone configuration* (Fig. 7b) using eq. (2). This projected representation entirely defines the flow as we consider it 2D on the surface of the cone, so as stated before $z = f(x,y)$ is completely constrained by the cone shape.

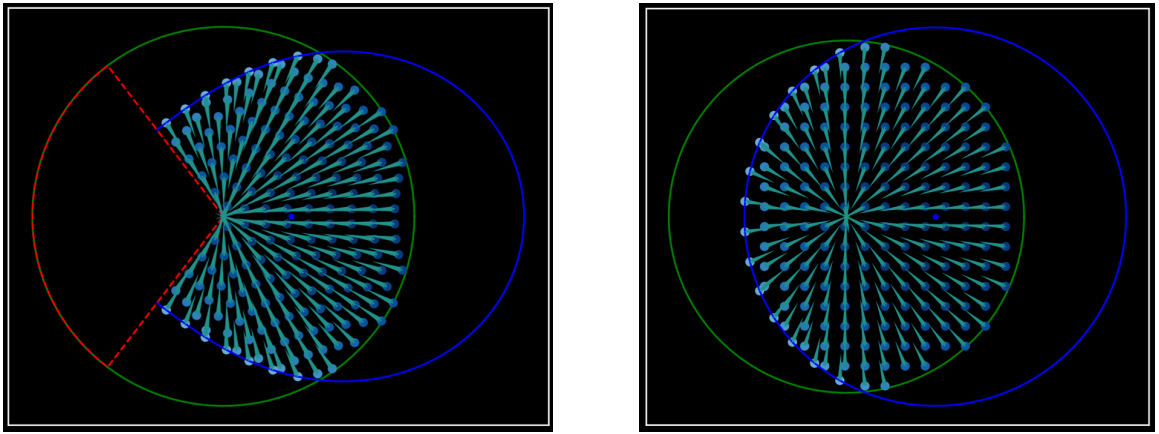

(a) Tangential velocity in *disk configuration*  (b) Tangential velocity in *cone configuration*

Figure 7: Tangential velocity in both configuration with $\frac{r}{R_d} = 0.5$.

We then compute the mean flow in this configuration and apply the mean velocity to all points (Fig. 8a). Finally we retranspose these velocities in the *disk configuration* (Fig. 8b) to get the advection velocity. This describe the fact that, if there was no drop deformation due to the normal impact, the average movement of the drop fraction would be to slide along the cone a the average tangential velocity.

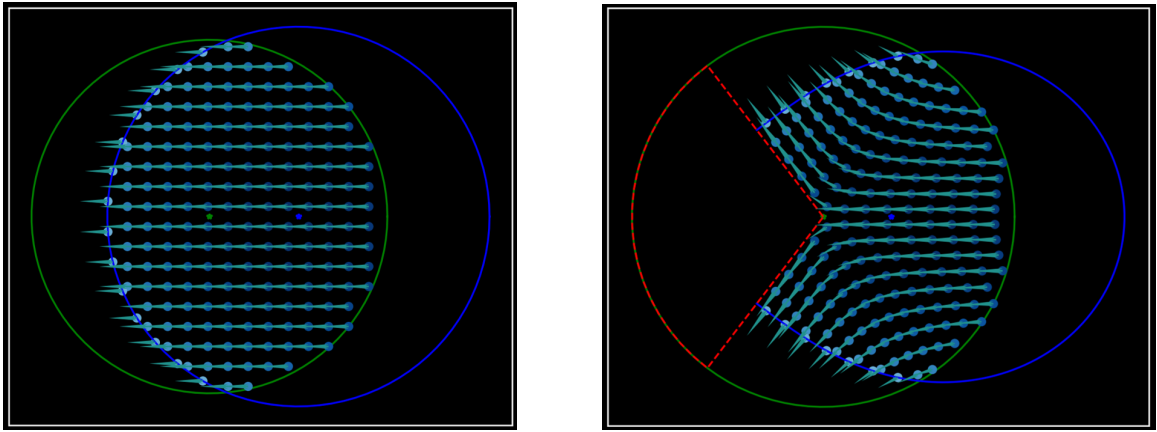

(a) Mean tangential velocity in *cone configuration*

(b) Mean tangential velocity in *disk configuration*

Figure 8: Mean tangential velocity in both configuration with $\frac{r}{R_d} = 0.5$.

**Full velocity field.** The total velocity for each impact point is the sum of the normal impact velocity and the tangential advection velocity (Fig. 9). As stated earlier this initial velocity is constant on the disk surface, in the *disk configuration*. The fluid particles trajectories will be straight until they contact either the cone border or the removed sector. In the next section we will discuss what happens then and describe the jet formation mechanism.

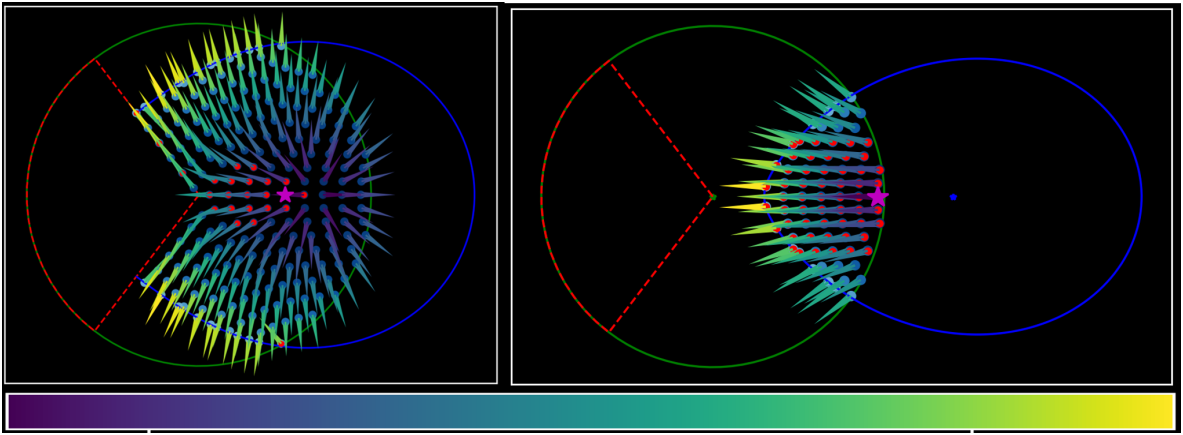

Figure 9: Total velocity for two cases. The left case is with $\frac{r}{R_d} = 0.5$ and the right case with $\frac{r}{R_d} = 1.4$.

### 2.1.5 Jet formation mechanism

As said before we hypothesize that the focusing of fluid trajectories by the cone's curvature is responsible for jet formation. We can observe on Fig. (2b, left) that some of the trajectories meets the border of the removed sector. If we consider a single trajectory, it should reappear on the other border with the same angle and keep going. Since we placed the removed sector at the opposite of the impact point, a second symmetrical trajectories also meets the sector border on the other side (see Fig. 2b, right). At the point of contact we consider that the momentum of the fluid particle following each trajectories are added, resulting in a radial velocity with a direction and amplitude depending on the angle of the contact (Fig.10-ABC). The fluid particles that reach the border of the cone are considered to be in ballistic trajectory which means that in the *cone configuration* their trajectories are rectilinear in the $(x, y)$ plane. By now working in this configuration, we can determine for each fluid particle if it joins the jet (by crossing the removed sector outside the cone) or is part of the fluid sheet (Fig.10-DE).

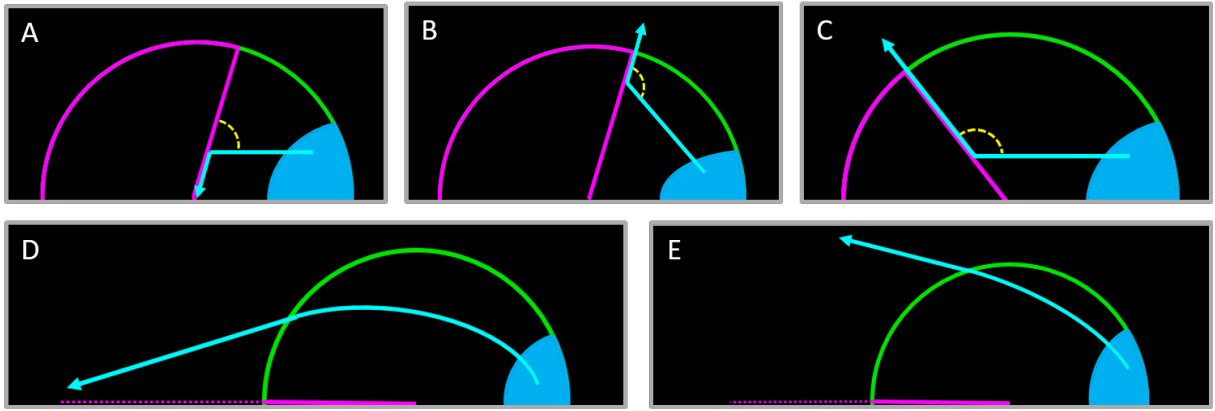

Figure 10: Schematics of the different possible fate for a given trajectory.

In experiments, a second type of jet that goes upwards can be observed for cone with very low angles. If we look at this cases in the model, we can observe that some trajectories a so curved that when they reach the border of the removed sector they are going towards the center of the cone. Figure 11 shows a simulation where with both types of jet. The out of plane dynamics of these jets is beyond the scope of this model, and we cannot predict the direction of ejection, that has been observed to be variable in experiments, but we will quantify separately the amount of fluid going into each jet.

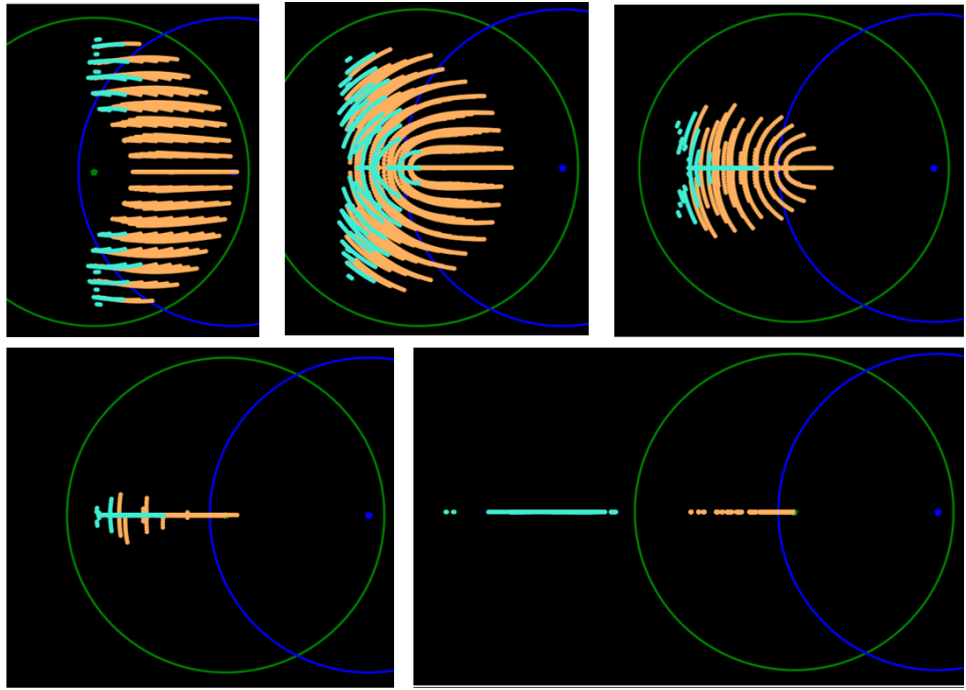

Figure 11: Illustration of lateral (green) and upward (orange) jets in a cone of 20° angle, for simulations. Time goes left to right and top down.

### 2.1.6 Evaluation of jet formation efficiency

**Impact volume fraction in the jet** We will quantify the jet volume fraction as the ratio of the volume of the jet relative to the total volume that impacted the cone. As introduced in section 2.1.5 we consider that the jet is composed of all the trajectories that touch the border of the removed sector and thus collide with each other. This can happen inside of the cone or after leaving it, and in the following we will give the geometrical conditions necessary to identify both types of trajectories.

**Colliding inside the cone** For the trajectories colliding inside the cone we can use the *disk configuration* and write the equations of the sector borders as:

$$y = C_{12}x \tag{7}$$

with

$$C_1 = \tan(\pi - \frac{\beta}{2}) \quad \& \quad C_2 = -C_1$$

For each trajectory starting at a point $M' : (x_{M'}, y_{M'})$ with an initial velocity $V'_{M'} : (V'_{x_{M'}}, V'_{y_{M'}})$ we can also write the equation of the trajectory as

$$y = Ax + B \tag{8}$$

with

$$A = \frac{V'_{y_{M'}}}{V'_{x_{M'}}} \quad \& \quad B = y_{M'} - \frac{V'_{y_{M'}} x_{M'}}{V'_{x_{M'}}}$$

The two lines intersect when (7) and (8) are equal. We get the coordinates of the intersection points with the two borders:

$$(x_{inter_{12}}, y_{inter_{12}}) = (\frac{B}{C_{12} - A}, \frac{BC_{12}}{C_{12} - A}) \tag{9}$$

For $\beta > 0$ and $V_{xy_M} \neq 0$, at least one of these two points exists for each trajectory. We then need to check if the intersection points are inside the cone (in *disk configuration*):

$$x_{inter_{12}}^2 + y_{inter_{12}}^2 \leq \frac{R_c^2}{\sin^2 \alpha} \tag{10}$$

First we will consider only the points from (9) verifying (10) and will deal later on with the trajectories joining the jet outside the cone. We then check if the remaining points are in the correct quadrant (intersection with the actual removed sector and not its symmetrical) and if the velocity is directed towards the sector. For cone angles $\alpha < 30°$ ($\beta > 180°$) points and velocities need to verify:

$$x_{inter_{12}} > 0$$

and

$$\text{sign}(y_{inter_{12}}) = \text{sign}(V'_{y_{M'}}) \tag{11}$$

For cone angles $\alpha > 30°$ ($\beta < 180°$) the conditions are:

$$x_{inter_{12}} < 0$$

and

$$\text{sign}(V'_{x_{M'}}) < 0 \tag{12}$$

Finally, for $\alpha < 30°$, we need to assign trajectories to both the lateral and upward jet. In order to do so we compute the scalar product of the velocity vector $(V_{x_M}, V_{y_M})$ with the vector defining the intersected border $(1, C_{12})$. If the product is positive the trajectory is heading into the lateral jet, otherwise it will be part of the upward jet.

**Colliding outside the cone** All the trajectories that do not meet in the cone will have an exit point defined by $(x_{exit}, y_{exit})$

$$x_{exit} = x_{M'} + \lambda V'_{x_{M'}}$$

and

$$y_{exit} = y_{M'} + \lambda V'_{y_{M'}} \tag{13}$$

with $\lambda > 0$ such that:

$$x_{exit}^2 + y_{exit}^2 = \frac{R_c^2}{\sin^2 \alpha} \tag{14}$$

Knowing the exit point $(x_{exit}, y_{exit})$ and the velocity $(V'_{x_{M'}}, V'_{y_{M'}})$ and in the *disk configuration* for each trajectory we can compute their counterpart $(X_{exit}, Y_{exit})$ and $(V_{X_{exit}}, V_{Y_{exit}})$ in the *cone configuration* using (1) and (2). Once outside of the cone the fluid particle have a balistic trajectory (in our purely kinematic model) that is a straight line when viewed from the top in the *cone configuration*. From the exit point and 'initial' velocities we just computed, we can discriminate between the trajectories joining the jet (y = 0 in the *cone configuration*) and the ones that participate in the fluid sheet. The trajectories joining the lateral jet outside verify in the *cone configuration*:

$$V_{X_{exit}} < 0$$

and

$$Y_{exit} . V_{Y_{exit}} < 0 \tag{15}$$

To get the final volume fraction in each of the jets, we identify the trajectories $t_M$ that form the jet ($I_{t_M}^{tan/vert} = 1$) and the ones that are not in the jet ($I_{t_M}^{tan/vert} = 0$). To get the volume fraction we compute the height of the column of water that impact each $t_M$'s starting point $M : (x_{t_M}, y_{t_M})$ in the *cone configuration* using the equation for a sphere centred in $(r, 0)$:

$$z_{t_M} = 2\sqrt{R_d^2 - (x_{t_M} - r)^2 - y_{t_M}^2} \tag{16}$$

The impact volume fraction in the jets is the average of the trajectories weighted by their height:

$$VolFracJet^{tan/vert} = \frac{\sum_{t_M} I_{t_M}^{tan/vert} z_{t_M}}{\sum_{t_M} z_{t_M}} \tag{17}$$

**Total volume in the jets** We can get to the total volume ejected in the jets by computing the fraction of the drop volume that impacts the cone. This volume is the intersection volume of an infinite cylinder of radius $R_c$ centred on $(0, 0)$ with a sphere of radius $R_d$ centred on $(r, 0)$. This complex geometry problem has been solved by [Boersma and Kamminga, 1961] and has been implemented in python using the `mpmath` python library for the computation of the elliptic integrals. We can get the impacting volume as

$$ImpactVolume = DropVolume \times ImpactFraction$$

and the jet volume as:

$$JetVolume = ImpactVolume \times VolFracJet$$

**Kinetic energy available for dispersal** The last quantifications we will define is the kinetic energy available for dispersal in each jet. We consider the kinetic energy associated with the volume in a jet:

$$E_{Jet} = \frac{1}{2}\rho \times JetVolume \times V_{eq}^2 \qquad (18)$$

where $V_{eq}^2 = \frac{\sum_{t_M} I_{t_M} z_{t_M} V_{t_M}^2}{\sum_{t_M} I_{t_M} z_{t_M}}$ with $V_{t_M}$ the velocity of the fluid particle trajectory originating in $M$ at the position where it enters the jet. For the lateral jet we take the velocities after the colliding of the two trajectories and for the upward jet we use the velocities of the particles before the impact as we do not know the direction taken by the jet out of plane.

This computation ignores a lot of the details of the jet formation mecanism and the dispersal of the incoming water. Here we consider the total kinetic energy that will 'feed' the jet over the whole process and do not account for the various losses and dissipation effects. It represents the 'maximum available' kinetic energy for each jet, and is reported relative to the total incoming kinetic energy:

$$E_{rel} = \frac{E_{Jet}}{E_{Drop}} = \frac{2E_{Jet}}{\rho \times \frac{4}{3}\pi R_{Drop}^3 \times V_{Drop}^2} \qquad (19)$$

## 2.2 Simulation details

We now have a complete model to simulate drop impacts on cones of various shapes and we defined several output variables to quantify the formation and efficiency of the jets.

### 2.2.1 Drop mesh

One important parameter of the simulations is the mesh of starting points for trajectories (Fig. 12). The mesh is defined in the *cone configuration* and then transposed in *disk configuration* using (1). We start from a square mesh of size $2R_d$ centred on $(r, 0)$ like the drop (Fig. 12a). The number of points in the mesh is $n^2$ with $n = 71$ for every simulation results shown. Then the mesh of impact points is constituted of the points of the square mesh that satisfy both (4) and (5) i.e. they are in the intersection of cone and drop (Fig. 12b). Additionnaly $2n$ points are positionned on the drop contour (Fig. 12a) and only those satisfying (4) are kept (Fig. 12b). For each of those points we use (16) to compute the height of the drop (Fig. 12c and 12d).

### 2.2.2 Parameter spaces

We must now define the parameters space that we want to explore with simulations in order to understand the shape optimisation of splash-cups. Two different types of simulations were run using the model: a single drop to compare to experiments and a random rain to infer the average optimal cup shape.

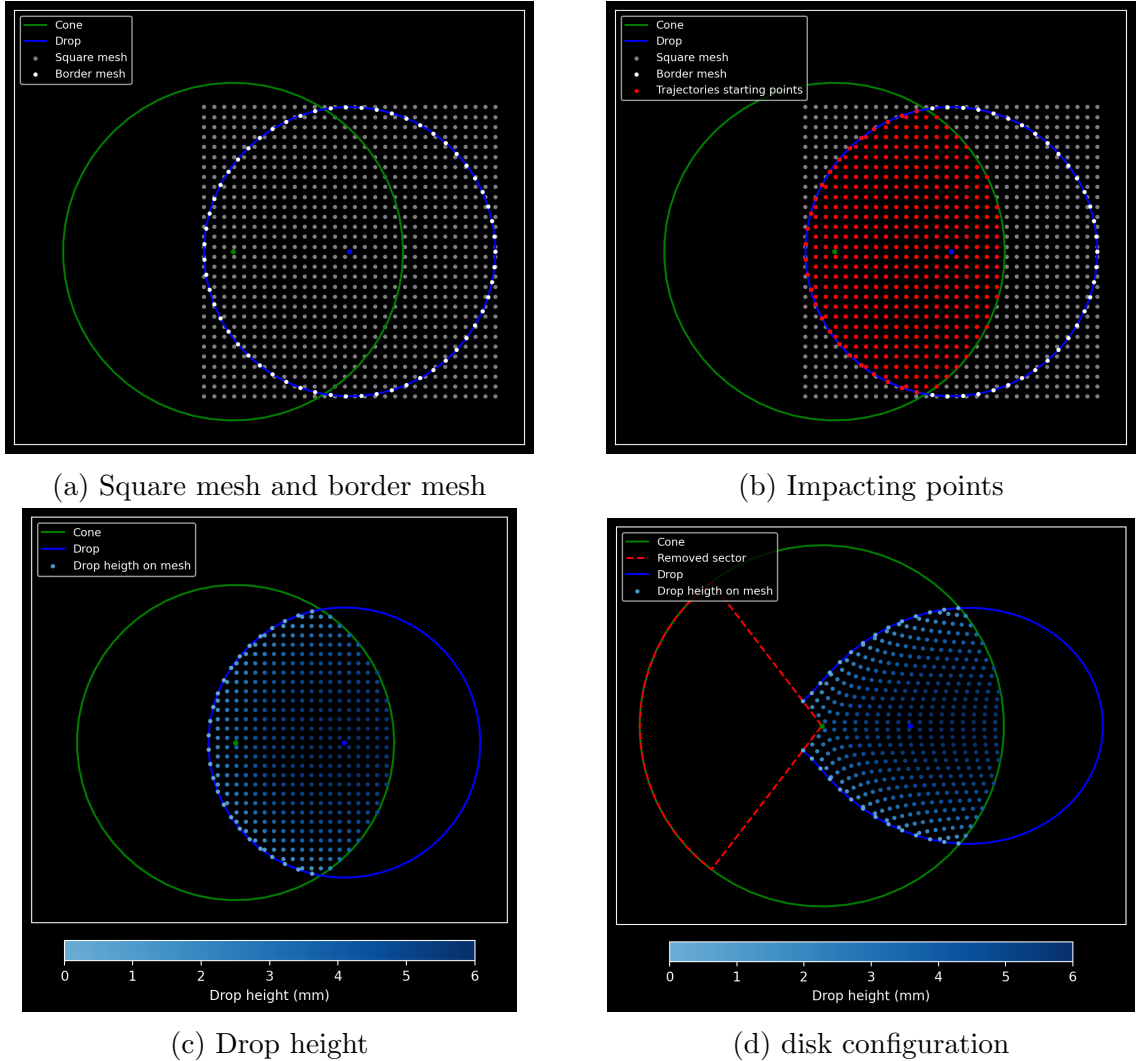

(a) Square mesh and border mesh  (b) Impacting points

(c) Drop height  (d) disk configuration

Figure 12: Mesh of trajectories starting points

**Single drop**  The relevant parameters of the problem are: the drop diameter $R_d$, the cone diameter $R_c$, the cone angle $\alpha$ and the distance between cone and drop center $r$. From these we naturally define a 3D parameter space:

1. The cone angle $\alpha \in [0, \frac{\pi}{2}]$

2. The relative drop size $RS = \frac{R_d}{R_c} \in [0, 2]$

3. The relative drop off-centring $OC = \frac{r}{R_c} \in [0, 3]$

Note that for $RS < OC$ the drop doesn't touch the cup. Additionnally, for $RS > OC + 1$ the drop covers the whole cone and we consider that there is no jet based on experimental observations.
This parameters space is a good one for evaluating separately the influence of each geometrical parameters, and is easy to compare with experimental data. One caveat is that the surface of the cone varies greatly with the angle for a fixed radius (Fig.13).

**Cone of constant surface**  While this is not a problem to analyse the jet formation mechanism and compare with experiments where the geometry is chosen, it makes little sens in the context of plant shape otpimisation. Indeed the amount of material that can be produced by the plant to form the cone is limited, and thus we decided to alter the parameters space to reflect this constraint. For the shape optimisation with random rain simulations the cone's surface (S) is fixed, and the radius is completely determined by the angle through the formula

$$R_{cone} = \sqrt{\frac{S \tan(\alpha)}{\pi}} \qquad (20)$$

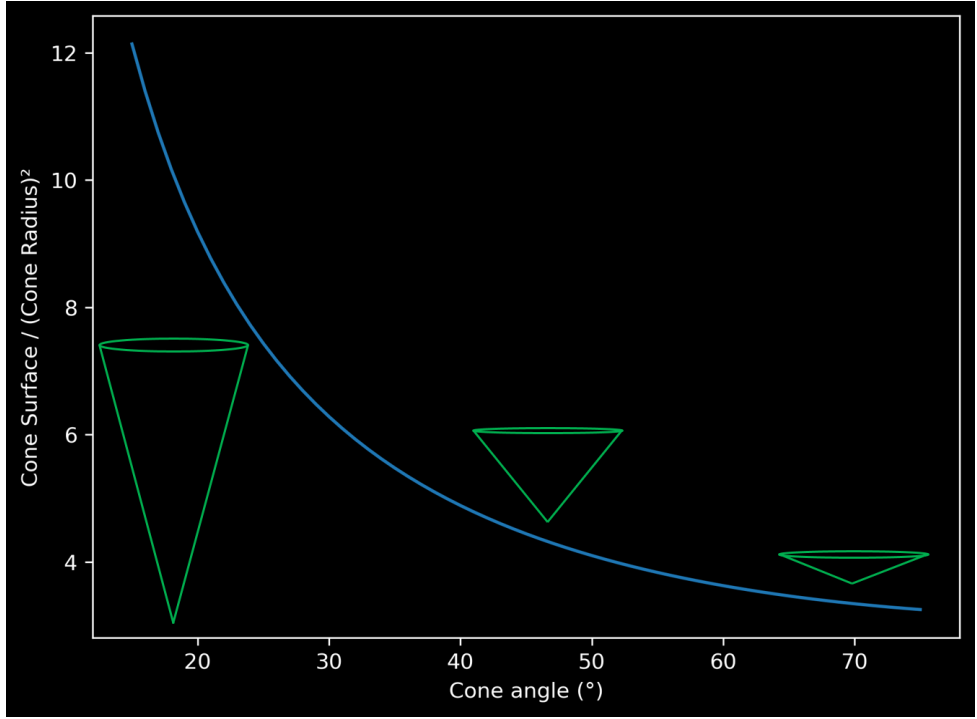

Figure 13: Evolution of the cone's surface with the angle, for a fixed radius

This additionnal constraint will have the effect of reducing the radius of small angle cones, which will be important in the context of rain, as a smaller cone catches less drops. Additionally in order to consider primarily the morphological parameters of the plant, we will also express the adimensionnal size of the cone as the cone surface divided by the average drop surface instead of the drop radius divided by the cone radius as was done in the single drop study.

**Random rain** In order to find the optimized shape for the plant cup, we want to place ourselves in a situation as close as possible to the biological context. For this we will simulate a rain of many drop impact on the same cone with different drop characteristics chosen to emulate a natural rain. For the raindrop size distribution we use a gamma distribution (Fig. 14a) and we take the median drop surface to normalize the cone surface. We compute the impact velocity as the terminal velocity in air depending on the drop size:

$$v(d)^2 = \frac{4}{3C_D} \frac{\rho}{\rho_{air}} gd \qquad (21)$$

with $C_D$ the drag coefficient, taken to be 0.5 assuming the drops stay spherical. Fig. 14b shows the velocities associated with the distribution of Fig. 14a. The off-centering of the drops is randomly chosen from a homogenous spatial (2D) distribution.

For each cone geometry (surface + angle) we simulate the impact of all the drops in the rain on it, and we measure the quantities previously defined. The jet fraction is the total volume ejected as a jet compared to the total impacting volume. The jet kinetic energy

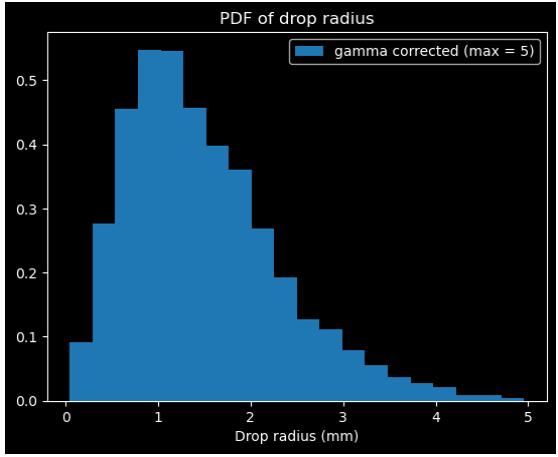
(a) Simulated rain drop size distribution

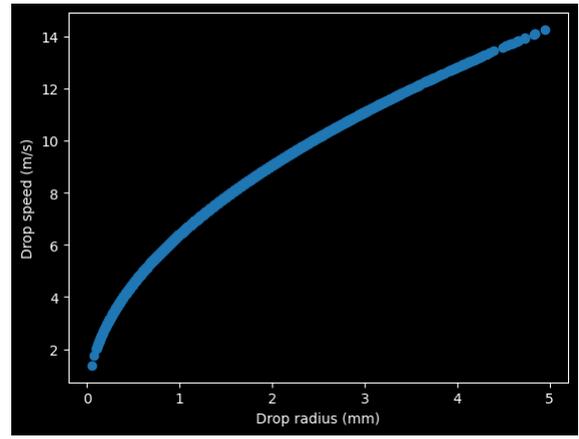
(b) Velocities associated with the drop sizes

Figure 14: Random rain characteristics

is the total kinetic energy of all the jets formed divided by the total kinetic of the rain. The dispersal distance of the lateral jet and height of the upward jet are expressed with the median as the average, since the distribution of dispersal distances do not have a fixed shape for different cones (see Fig.15).

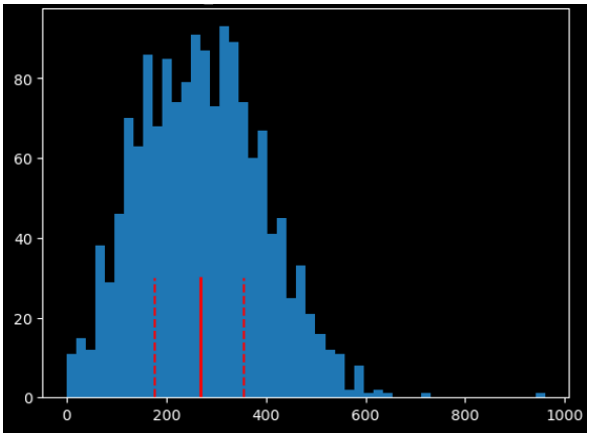
(a) Dispersal distance distribution for a cone of 34° angle

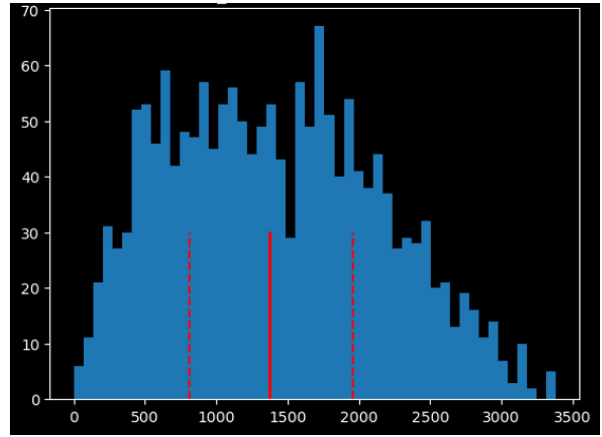
(b) Dispersal distance distribution for a cone of 45° angle

Figure 15: Example of dispersal distance ditribution for different cone shapes. The red line is the median, the red dashed lines are first and last data quartile.